# Drive for Creativity

Leonid Perlovsky[a] and Dan Levine[b]

[a]Harvard University and AFRL leonid@seas.harvard.edu,
[b]Department of Psychology, University of Texas at Arlington, levine@uta.edu

*Abstract*—**We advance a hypothesis that creativity has evolved with evolution of internal representations, possibly from amniotes to primates, and further in human cultural evolution. Representations separated sensing from acting and gave "internal room" for creativity. To see (or perform any sensing), creatures with internal representations had to modify these representations to fit sensor signals. Therefore the knowledge instinct, KI, the drive to fit representations to the world, had to evolve along with internal representations. Until primates, it remained simple, without language internal representations could not evolve from perceptions to abstract representations, and abstract thoughts were not possible. We consider creative vs. non-creative decision making, and compare KI with Kahneman-Tversky's heuristic thinking. We identify higher, conscious levels of KI with the drive for creativity (DC) and discuss the roles of language and music, brain mechanisms involved, and experimental directions for testing the advanced hypotheses.**

## I. INTRODUCTION

PRIMITIVE creatures, such as lobsters, do not use extensive internal representations in the mind. Their neurons are "hardwired" for relatively few crudely defined situations and sensor signals almost directly wired to muscles. Although the intermediate neural circuits in invertebrates between sensory organs and muscles are plastic, the learning and adaptation that takes place amounts to forming simple stimulus-response connections that do not generalize to other situations. There is no creativity. With evolution of adaptive representations, the variety of objects and situations in the world that the mind can recognize grows tremendously. Adaptive representations imply that yesterday's representations may not fit today's reality, and learning mechanisms are necessary for adaptation. Moreover, adaptation becomes necessary, because without adaptation perception becomes impossible. Therefore the *knowledge instinct* (KI), that is, the drive to adapt mental representations to the surrounding reality, had to evolve along with adaptivity. We discuss this mechanism, the mathematical reasons and experimental evidence that learning proceeds from vague to crisp representations [1-5].

Without language, adaptivity is limited to what can be perceived by sensors and to a limited number of combinations of perceptions. Visual and other sensory cortex circuitries in higher animals are tremendously complex mechanisms. Still we suggest that animals cannot learn to generalize much beyond direct perceptions; consequently complex animal behavior has to be based on inborn mechanisms with minimal adaptations. We discuss the reason for this hypothesis and how it can be tested.

With language, cultural evolution outpaced genetic evolution, knowledge accumulated fast, and a hierarchy of abstract representations evolved. Abstract thoughts in language are grounded in external society and kids by age 5 can talk about almost everything ([6], [7]). Learning language is guided by a variant of KI (the Language Instinct, LI; [8]), and proceeds from vague to crisp representations (including syntax and complex abstract thoughts). Whereas KI fits internal cognitive representations to sensor perceptions, LI fits internal representations of language (vague at birth) to external language. But LI does not connect language to the world. We discuss a mechanism connecting language to cognitive representations, so that matching cognitive representations to the world by KI is guided by language ([6], [7], [9]). This hypothesis is fundamental to understanding creativity, by means of understanding the difference between creative and heuristic thinking. We discuss its theoretical bases, experimental evidence, and future tests [10].

We emphasize that creativity is a matter of degree. Autonomous and unconscious working of KI usually is not called creativity. Conscious mental effort resulting in new understanding-representations in the mind of a single person is a creative process, even if this understanding is not new for a society; this is the distinction Boden [11] and others draw between *personal creativity* and *historical creativity*. Every normal kid understands surrounding language by the age of 5; these mechanisms are autonomous and non-creative. Yet even at 5 sometimes kids use language creatively: could this be studied scientifically? Reading and writing are not autonomously acquired skills. Skills acquired at an average level, with much tutoring and rote learning, usually are not called creativity. Should a better memory, quicker reactions, or faster speaking be called more creative? We emphasize the creative aspect of connecting language with cognition.

Since the time of Aristotle [12], the ability to speak well has been identified with intelligence. Yet some people can be cognitively creative while possessing only average ability for expressing their thoughts in language. Of course the two sides of SAT tests acknowledge this. We discuss scientific approaches to studying this phenomenon. Are the two parts of SAT adequate for testing creativity? What can be deduced from contemporary models of the mind?

The mind architecture has hierarchical structure, from elementary percepts to abstract thoughts. We suggest that creativity is manifest at higher levels. As mentioned, language is acquired early in life crisply and consciously at many hierarchical levels including the abstract ones, but we discuss why cognition at higher levels may remain fuzzy-vague. We discuss that crispness of language masks vagueness of cognition from our consciousness. Subjectively, ideas may be perceived as consciously understood, if we can talk about them. The paper discusses a mechanism by which crispness of language hides from the consciousness the vagueness of corresponding cognitive representations. It is obvious when watching kids: they talk well, but they do not "really" know what it means. In regards to highly abstract ideas, most of the time most adults are like kids. Therefore it takes more creativity to make crisp and conscious cognitive model-representations at higher levels. This hypothesis can be tested experimentally.

Creativity in a wide sense develops crisp cognitive representations corresponding to crispness of language, in other words, to understand the world and self at the level existing in language and culture. In a more narrow and refined sense creativity has several aspects. Scientific creativity develops cognitive representations, which surpass language in their adequacy to the world (and self). Writers formulate knowledge in language crisper than it has been formulated previously; new knowledge is created in this process if stronger connections to cognitive representations are achieved. As we discuss that requires connecting language with emotions. In this area, writers, poets, and composers use various means for similar goals: emotionally connecting language and cognition.

Creativity, thus, involves not only conceptual knowledge but emotions as well. We discuss the role of emotions in cognition and identify specific emotions related to the KI and creativity ([2-4], [6], [13-17]). Higher up in the hierarchy of the mind, emotions and concepts are intermixed in a non-differentiated fashion. Development of differentiated knowledge, especially at higher levels of the hierarchy, creates psychological tensions, that is, cognitive dissonances. These dissonances in the psyche interfere with unity, which is necessary for achieving higher goals, psychic health and sometimes even survival. To continue developing knowledge and to be able to cope with this multitude of dissonances, humans need the corresponding multitude of conscious "higher" emotions that help one to bring the "feelings" of the dissonances into consciousness, and then to cope with them. Poetry and music evolved for this purpose [18-20]. Closely related is a need to continue connecting language to cognition, which requires this same multitude of emotions ([7], [15]). This is the cognitive role of poetry, songs, and musical theater. Correspondingly, there are emotional creativities of poets, composers, musicians, singers and also writers – they create new emotions necessary for being able to cope with the diversity of cultures – to connect language with cognition, and to continue the process of cultural evolution [15].

Cognitive representations include understanding and behavior. We discuss that the highest conceptual understanding relates to the beautiful. The highest of behavioral representations relate to the spiritually sublime [10].

The next section discusses mechanisms of KI, a contradiction between knowledge-maximizing mechanisms of creativity driving cultural evolution of human highest spiritual abilities on one hand and, on the other, heuristic mechanisms minimizing cognitive-effort as revealed by Tversky and Kahneman [21, 22]. Then we discuss mechanisms of interaction between cognition and language and relate them to the contradiction between KI and heuristics. The discussion section addresses experimental approaches to demonstrating the difference between heuristic and KI mechanisms, the role of emotions in creativity, and experimental approaches to demonstrating drive for creativity [17].

How far toward understanding creativity can we move based on contemporary models of the mind?

## II. HEURISTICS VS. THE KNOWLEDGE INSTINCT

Research initiated by Tversky and Kahneman [21, 22] demonstrated that decision making is often irrational and based on heuristics, which may or may not be suitable to a particular situation. Humans often prefer fast decisions, saving effort, including mental effort, to original thinking. According to Maimonides [23], refusing original thinking and choosing a shortcut to the ready-made knowledge was the reason for expelling Adam from paradise. Maimonides' explanation was connected to contemporary cognitive science in [10].

Fast decisions saving time and effort have clear evolutionary advantages. Efficient use of heuristics could be counted as "personal creativity." But if irrationality and heuristics are fundamental to human decision making, what is the source of all human knowledge? Heuristics might be used well or poorly, but what is the source of heuristics? This source should be no less fundamental to human psyche than irrationality. We suggest that heuristics is a store of accumulated cultural knowledge, whereas the source of this knowledge is the KI. Heuristics originate in those relatively rare times when creative individuals use their KI and create new knowledge that turn useful for the rest of the society (creativity or "historical creativity").

Perlovsky and McManus [24], (see also [1] and [2]), developed a mathematical theory of the knowledge instinct (KI), the drive to make sense out of one's environment. Their mathematical description of the KI involves maximizing similarity between mental representations-models and patterns in sensory signals. More generally, KI maximizes similarity between bottom-up and top-down signals, which is interpreted as knowledge. Top-down signals produced by mental representations, indexed by $m$ at each hierarchical level. In neural terminology they prime receiving neurons, $n$, to recognize pattern $\mathbf{M}_m(n)$ in the bottom-up signals $\mathbf{X}(n)$. Mathematically, they predict the expected patterns in the bottom-up signals; the similarity between bottom-up signals $\mathbf{X}(n)$ and top-down signals $\mathbf{M}_m(n)$ ([1], [2]) is expressed as

$$L(\{\mathbf{X}\},\{\mathbf{M}\}) = \prod_{n \in N} \sum_{m \in M} r(m)\,\ell(\mathbf{X}(n) \mid \mathbf{M}_m(n)). \quad (1)$$

This similarity expression considers each top-level representation, m, as an alternative, a possibility of an object or situation m to be present among bottom-up signals n. Conditional similarities

$\ell(\mathbf{X}(n) \mid \mathbf{M}_m(n))$ can be modeled [25, 26] as

$$\ell(\mathbf{X}(n) \mid \mathbf{M}_m(n)) = \prod_{i=1}^{D} p_{mi}^{xni}(1 - p_{mi})^{(1-xni)} \quad .(2)$$

Here, n indexes bottom-up signals, m indexes top-down signals, and i indexes their constituent components; e.g. features of a pattern, or objects making up a situation, or situations making up an abstract representations. A bottom-up signal $x_{ni}$ indicates a presence of a component i in a bottom-up signal $n$, and a top-down signal $p_{mi}$ models a probability of a component $i$ in a top-down signal $m$. Thus expressions (1) and (2) model similarity measures at every level of the mind hierarchy. The number of possible components, $i$, is denoted as D. All possible associations between bottom-up signals, N, and top-down signals, M, are accounted for in similarity ($l$). This is a huge combinatorial number on the order of $N^M$.

As discussed in [1] and [2], this is the reason for the computational complexity encountered by most artificial intelligence, pattern recognition, and neural network algorithms since the 1950s. Such complexity is related to classical logic [27, 28], which is a part not only of logical rule systems, but also of those approaches specifically designed to overcome logical limitations. Classical logic is used in training procedures of pattern recognition and neural network algorithms, and in fuzzy systems' selection of the degrees of fuzziness. This is the reason for the seemingly irresolvable computational complexity of most algorithms encountered for decades. The above references describe Neural Modeling Fields (NMF), a mathematical procedure that uses dynamic logic (DL) to overcome limitations of computational complexity. Resulting algorithms outperform previous approaches by orders of magnitude ([29-33]). Whereas classical logic involves crisp statements and thus is a static statement-logic, DL is a process-logic; it involves evolution of mental representations from vague-fuzzy to crisp. It describes emergence of approximately classical logic processes from illogical and fuzzy neural processes in the brain ([7], [35]). Neuroimaging experiments demonstrated that visual perceptions in the human brain follow predictions of NMF-DL ([1], [2], [5], [7], [16]). Thus at the basic level of visual perception, DL models a mechanism of KI.

At lower levels of the brain hierarchy KI operates autonomously, serving as a foundation of perception. However, our hypothesis in this paper is that at higher levels KI operations are not autonomous, they require conscious effort, and they serve as foundations for creativity. KI at higher levels becomes a drive for creativity. This is because existing concepts or combinations of concepts are often inadequate for resolving cognitive and emotional dissonances between top-down and bottom-up signals between different high levels. The roles of several prefrontal and subcortical regions in forming, breaking, and changing behavioral and decision rules are discussed in [34].

At lower levels of perception there is no competition among mechanisms of decision making. Perception cannot operate heuristically; higher animals as well as human beings have to use DL mechanisms of KI to be able to orient in the surrounding world. In the next section we discuss decision making at higher cognitive levels and relate a competition between KI and Kahneman-Tversky heuristic mechanisms to a competition between cognitive and language mechanisms.

## III. KI AND HEURISTICS VS. COGNITION AND LANGUAGE

Let us repeat that at lower levels of perception KI operates autonomously. This is similar in animals and in humans. Here we relate ability for higher level abstract thinking to language, and in turn, relate language to a competition between KI and heuristic thinking.

As argued in [15] and [36], higher level abstract thinking is only possible due to language. To model this ability we have to consider mechanisms of interaction between cognition and language, and their parallel hierarchy, Fig. 1 ([2], [6], [14], [31]). NMF-DL describes language similar to cognition by the hierarchy of levels maximizing similarity (1). A newborn human mind does not have specific models for concrete words or syntactic rules. Instead, it contains vague "placeholders" that in the process of learning would acquire specific contents (words, rules) matching the language spoken around. Similarly, a newborn human mind does not have specific models for concrete objects or situations; it contains vague "placeholders" that in the process of learning would acquire specific contents (images, relations) matching surrounding world. Interaction between cognition and language is modeled mathematically by the mechanism of dual models. While language and cognitive models in the newborn brain are vague placeholders, the neural connections between these future models are inborn ([6], [15]). Therefore human brain-mind does not have to "figure out" which words go with which objects (this mathematically unsolvable problem can be solved using the dual model, [37-41]). As vague placeholders acquire concrete contents, connections between words and objects, phrases and situations are already in place. This is the mechanism of the dual model, which enables language-cognition interaction.

Note that as Fig. 1 indicates, language is acquired from surrounding language ready-made at all hierarchical levels. This is the reason kids of 5 years old can talk about complex abstract ideas, which they do not yet have experience to "really understand." But note also that cognitive experience is acquired from the surrounding world only at the very bottom of the hierarchy, at the level of sensory perceptions. Learning at this level (perception) is possible without language, and animals can perceive objects, and to some degree situations. But learning abstract ideas, which are combinations of elements of many situations, combinations that cannot be directly perceived, is impossible from direct experience. The reason is that the number of combinations is too large, much larger than any individual experience will enable to learn useful abstract ideas from random sets of objects and relations. Only due to guidance from language, abstract cognitive models could acquire concrete contents. KI drives every human brain-mind to develop abstract cognitive contents corresponding to language models.

Higher up in the hierarchy cognitive models remain vaguer and less concrete. Learning-understanding and creating new models continue throughout life. Higher in the hierarchy the autonomous KI gradually turns into drive for creativity. It becomes less autonomous and more dependent on individual conscious effort. This is why there are significant differences between people in how well they "really understand" the contents of the spoken language. Our hypothesis is that one "really understands" when his or her cognitive models become crisp and specific corresponding to crispness of language. This hypothesis can be experimentally tested. We would emphasize that "real understanding" requires more than just crispness and concreteness. An abstract idea involves more than one crisp and concrete content; the drive for creativity strives for multiple differentiated contents, which have to be properly related and unified.

It follows from the above discussion that the drive for creativity has an inborn autonomous component; however, this inborn component weakens at the higher levels. Language, while providing guidance to the development of abstract cognitive models, also creates impediments to this process. To understand the nature of this difficulty, let us consider cognition at the lower level of perception, where it does not involve language. Close your eyes for 2 seconds and try to imagine an object that you have clearly seen in front of your eyes. The imagined object is vague, not as crisp and clear and not as consciously perceived as the object with opened eyes. Bar et al. [5] performed a similar and much more detailed experiment using brain imaging. The imagined vague object is created on the visual cortex by projections of mental representations, top-down signals. When you open your eyes, the object again is perceived crisp and clear, and it is impossible to perceive the vague imagined object with opened eyes. With opened eyes, crisp and consciously perceived projections from the retina, which are created by interacting bottom-up and top-down signals, mask less conscious and vaguer top-down signals. Similar processes occur at higher level. Language models act as "eyes" for cognition of abstract concepts. Because language models are crisp and conscious, they mask vague and unconscious abstract cognitive models. Therefore it is difficult to perceive that cognition is vague and inadequate. Every step towards more crisp and conscious cognitive abstract models requires conscious creative effort. Still, evolution of human knowledge and culture confirms that drive for creativity makes this development possible.

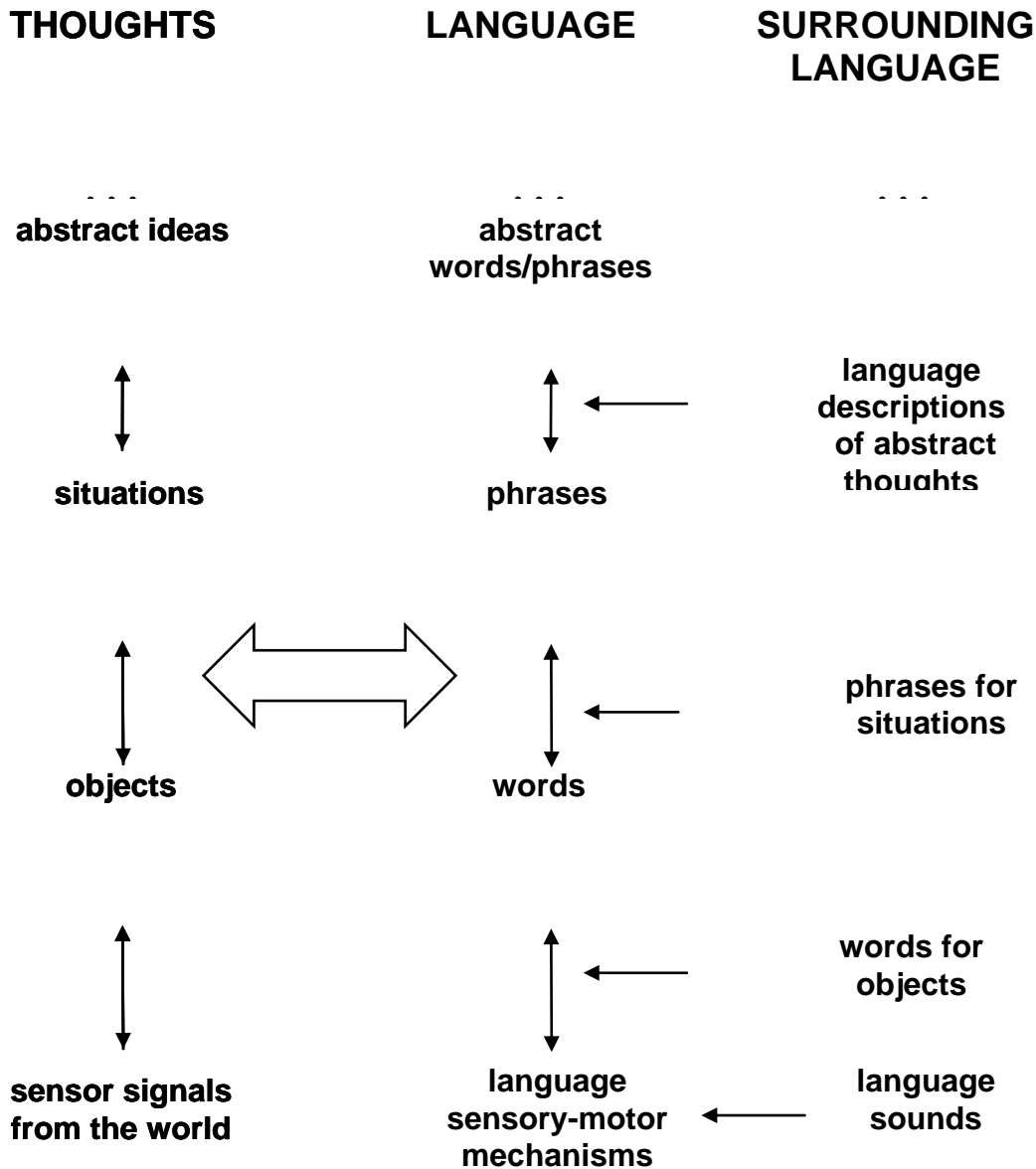

Fig. 1. Hierarchical integrated language-cognition MF system. At each level in a hierarchy there are similarities, models, and mechanisms of learning of language and cognitive models ([15]). The dual models (thick arrow) integrate language and cognition. Mathematically, similarities are integrated as products of language and cognition similarities. Initial models are fuzzy placeholders, so integration of language and cognition is sub-conscious. Association variables depend on both language and cognitive models and signals. Therefore language model learning helps cognitive model learning and v.v. High-layer abstract cognitive concepts are grounded in abstract language concepts similarities.

Now we can better understand the interplay between drive for creativity and irrationality of Kahneman-Tversky heuristics. Drive for creativity makes possible the improvement of abstract cognitive models. Yet, it takes much effort to overcome heuristic knowledge accumulated in language. Irrationality and heuristic thinking, according to the hypothesis of this paper, correspond to thinking with crisp language models, while the corresponding cognitive models remain vaguer and less conscious.

## IV. DISCUSSION

Steps toward mathematical modeling of interacting heuristic thinking and cognitive drive have been taken in [10] and [42]. Based on various fMRI results, we suggested in [10] that operations of heuristic mechanisms tend to activate the amygdala, a part of the brain that has changed little from other mammals to humans. On the other hand, creative operations of the knowledge instinct tend to activate executive regions of prefrontal cortex, a part of the brain whose complexity has grown exponentially in humans. Yet the roles of those regions are complex because connections between the amygdala and two regions of prefrontal cortex (orbitofrontal and anterior cingulate) are also implicated in higher-level emotional responses based on actual or expected consequences of stimuli or actions (see [42] for review). Hence those same connections likely play a role in the aesthetic emotional responses that accompany the intuitive elements of creativity.

Experimental tests of this hypothesis are ongoing. Here we continue developing this hypothesis by suggesting specific roles of cognition and language. We suggest that thinking by using ready-made heuristics involves language brain centers to larger degrees than original thinking using cognitive mechanisms. The connection between heuristics and language could be mediated by neural connections between cortical language centers and amygdala (either direct or through the thalamus). Whereas heuristics involve vague and less conscious cognitive representations, original thinking involves crisper and more conscious cognitive representations. This hypothesis can be experimentally tested using procedures similar to those used in referenced literature.

According to the theory of drives and emotions [13], every drive involves specific emotional neural signals indicating satisfaction or dissatisfaction of this drive. Specific emotions associated with KI were identified as aesthetic emotions ([4], [6], [7], [16], [18-20], [43], [44]). Experimental confirmation of aesthetic emotions related to satisfaction of KI was documented in [17]. This hypothesis has been further developed with our theory of differentiated KI, which in this paper we identify with the drive for creativity. It has been suggested that all inconsistencies in the body of knowledge, as well as all contradictions between knowledge and other instincts, create cognitive dissonances which are perceived emotionally as a multiplicity of aesthetic emotions. Maintaining unity of the psyche along with differentiated knowledge requires highly developed differentiated emotions, which are created in individuals and cultures by music ([6], [18-20]). The related form of creativity is a province of poets, composers, and singers. Affinity between emotions of cognitive dissonance and musical emotions is being tested by several research groups of psychologists and cognitive musicologists. This is a wide field for future research.

Also, an important part of the drive for creativity is the sense that we have achieved an understanding through our own efforts. Why should satisfaction of the KI through generating an idea on one's own provide us with a particular pleasure that we do not obtain as much from leaning about the same idea from someone else, or from a book or web site? The answer to this question probably involves a greater understanding of the nature of the self and how we differentiate the self from others. Aside from a general sense that the self is closely involved with prefrontal cortex, and that different and adjacent prefrontal regions are implicated in emotional and cognitive parts of the sense of self [43], the neural basis for the self-other distinction is largely unknown and is another field for future research.

The theory developed in this paper and others advances hypotheses about the nature of the highest aesthetic emotions of the beautiful and sublime ([2-4], [6], [7], [16-20], [44-47]. In the mind hierarchy, as shown in Fig. 1, concepts-representations at every higher level have evolved in cultural evolution for the purpose of creating a unified, general meaning of lower level concepts. Mental representations at the highest levels of the hierarchy strive for creating a unity of the entire human experience. While language representations at these highest levels have been developed by artists, philosophers, and theologians over millennia, cognitive representations remain vague and unconscious. Every step toward creating crisper contents of these representations improves understanding of the meaning of the entire life experience in its unity and is felt as emotions of the beautiful. Improving representations of behavior realizing the beautiful in one's life is felt as emotions of the sublime. Experimental tests of this hypothesis are a subject of future research.